\documentclass[twocolumn,aps,prl,showpacs,amsmath,superscriptaddress]{revtex4}
\usepackage{amssymb}
\usepackage{mathrsfs}
\usepackage{graphicx}
\usepackage{float}
\usepackage{amsmath}
\usepackage{color}

\begin{document}
\title{Dynamical topological invariant after a quantum quench}
\author{Chao Yang}
\affiliation{Beijing National Laboratory for Condensed
Matter Physics, Institute of Physics, Chinese Academy of Sciences, Beijing 100190,
China}
\affiliation{School of Physical Sciences, University of Chinese Academy of Sciences, Beijing, 100049, China}

\author{Linhu Li}
\affiliation{Department of Physics, National University of Singapore, 117542, Singapore}

\author{Shu Chen}
\thanks{Corresponding author: schen@iphy.ac.cn}
\affiliation{Beijing National Laboratory for Condensed
Matter Physics, Institute of Physics, Chinese Academy of Sciences, Beijing 100190,
China}
\affiliation{School of Physical Sciences, University of Chinese Academy of Sciences, Beijing, 100049, China}
\affiliation{Collaborative Innovation Center of Quantum Matter, Beijing, China}
\date{ \today}

\begin{abstract}
We show how to define a dynamical topological invariant for general one-dimensional topological systems after a quantum quench. Focusing on two-band topological insulators, we demonstrate that the reduced momentum-time manifold can be viewed as a series of submanifold $S^2$, and thus we are able to define a dynamical topological invariant on each of the sphere. We also unveil the intrinsic relation between the dynamical topological invariant and the difference of topological invariant of the initial and final static Hamiltonian. By considering some concrete examples, we illustrate the calculation of the dynamical topological invariant and its geometrical meaning explicitly.
\end{abstract}


\maketitle
{\it Introduction.-}
In the last decade, the study of topological quantum matter is one of the most attractive topic in condensed matter physics \cite{KAN,ZSC,TD,EW,SRY},
and our knowledge of topological properties for various quantum systems has been widely expanded. In contrast to equilibrium systems, what we know about the topological quantum matter out of equilibrium is quite rare \cite{CG}. The topology properties far from equilibrium have been studied in different ways, such as the dynamics of edge states \cite{MAM,AD,PDS}, dynamical quantum phase transition \cite{BD,MH,AVB}, Floquet topological states \cite{VG,PZ,GP}, etc. The rapid development of cold atom experiments provides a powerful tool to study the dynamics far from equilibrium \cite{IB,TE,IB2,PZ2}, and the evolution of a quantum state can be visualized with the method of Bloch state tomography \cite{CW,JJG,AE}.

A typical example of dynamics far from equilibrium is the quantum quench. Initially, the state is prepared in the ground state of Hamiltonian $H^i$, then we carry out a quench to the system by
suddenly changing a physical parameter, denoted by a new Hamiltonian $H^f$.
It is known that the topological invariant will remain unchanged because the time evolution operator is unitary \cite{MJB,MR,ZH}, and the topology of final Hamiltonian does not influence the topology of the time evolved state. However, the non-equilibrium topological response is found to exhibit some novel properties with no equilibrium analog \cite{YH,JHW,FNU,PW}.  Also, it has recently been shown
that the topology of final Hamiltonian can be reflected by the Hopf invariant in a two-dimensional (2D) Chern insulator \cite{ZH,JY,MT}, which gives a first example in defining the topological invariant far from equilibrium. It is still a challenge for understanding how to explore nontrivial topology properties of a dynamic system in a general way.

In this work, we study general two-band nonequilibrium systems in one dimension (1D), and extract a dynamical topological invariant defined on the momentum-time manifold. The 2D torus $T^2$ composed of momentum and time can be reduced into a series of spheres ($S^2$), and a dynamical Chern number is achieved from each of the spheres. This Chern number measures how many times that the Bloch sphere is covered when the time evolved Bloch vector winds over the corresponding sphere.
We also analyze the intrinsic relation between the dynamical topological invariant and the topology of $H^i$ and $H^f$ in equilibrium and give some examples to explain our results. At last, we point out that the dynamical topological invariant can be visualized experimentally.

{\it Model and quench dynamics.-}
We consider a general 1D two-band tight-binding model, and at each momentum $k$ the Hamiltonian is described by
\begin{equation}\label{Eq1}
h(k)=d_0(k)\mathbf{I}+\mathbf{d}(k)\cdot\boldsymbol{\sigma},
\end{equation}
where $\mathbf{I}$ is the $2\times2$ identity matrix and $\boldsymbol{\sigma}=(\sigma_x,\sigma_y,\sigma_z)$ are Pauli matrices acting on a (pseudo) spin-1/2 space. This model can be used to describe a variety of topological insulators and superconductors, for example, the Su-Schrieffer-Hegger model \cite{SSH} and p-wave Kitaev chain \cite{AYK}. The eigenvalues are given by
\begin{equation}\label{Eq2}
\epsilon_{\pm}(k)=d_0(k)\pm|\mathbf{d}(k)|,
\end{equation}
and we denote the eigenvectors as $|\psi_{\pm}(k)\rangle$. For convenience, we use the corresponding density matrices instead of the state vectors, which reads
\begin{equation}\label{Eq3}
\hat{\rho}_{\pm}(k)=|\psi_{\pm}(k)\rangle\langle\psi_{\pm}(k)|=\frac{1}{2}[1\pm\hat{\mathbf{d}}(k)\cdot\boldsymbol{\sigma}],
\end{equation}
where $\hat{\mathbf{d}}(k)=\frac{\mathbf{d}(k)}{|\mathbf{d}(k)|}$ is the normalized vector localized on the Bloch sphere $S^2$. The topological invariant of the system can be calculated with the information of $\hat{\mathbf{d}}(k)$ for both $\mathbb{Z}$ type and $\mathbb{Z}_2$ type in one dimension \cite{RMP1}.

Now we study the dynamical properties of the system far from equilibrium. By preparing the system in the ground state of the initial Hamiltonian $h^i$, i.e., $\rho^i(k)=\frac{1}{2}[1-\hat{\mathbf{d}}^i(k)\cdot\boldsymbol{\sigma}]$,
and then performing a sudden quench to the final Hamiltonian $h^f$, the evolution of density matrices can be written as
\begin{equation}\label{Eq4}
\rho(k,t)=\frac{1}{2}[1-\hat{\boldsymbol{d}}(k,t)\cdot\boldsymbol{\sigma}],
\end{equation}
with $\hat{\mathbf{d}}(k,t)$ achieved from the Liouville-von Neumann equation,
\begin{align}\label{Eq5}
\hat{\mathbf{d}}(k,t) =& \hat{\mathbf{d}}^i\cos(2|\mathbf{d}^f| t) + 2\hat{\mathbf{d}}^f(\hat{\mathbf{d}}^i\cdot\hat{\mathbf{d}}^f)\sin^2(|\mathbf{d}^f| t) \nonumber\\ & + \hat{\mathbf{d}}^i\times\hat{\mathbf{d}}^f\sin(2|\mathbf{d}^f| t),
\end{align}
where $\hat{\mathbf{d}}^i$ and $\hat{\mathbf{d}}^f$ are Bloch vectors of initial and final Hamiltonians \cite{SUP}, both of them are functions of momentum $k$. 
Eq. (\ref{Eq5}) can be interpreted as the winding of $\hat{\mathbf{d}}$ from the initial vector  $\hat{\mathbf{d}}^i$ around the axis $\hat{\mathbf{d}}^f$ on the Bloch sphere during the evolution process. The Berry phase of the time dependent wave function does not change after taking a global quench due to the evolution is unitary \cite{MJB}, hence it can not characterize the topological difference between $H^i$ and $H^f$. In the following text we shall explore how to define a unique topological quantity to characterize topologically different quench dynamics.

\begin{figure}[htbp]
  \centering
  \includegraphics[width=0.7\linewidth]{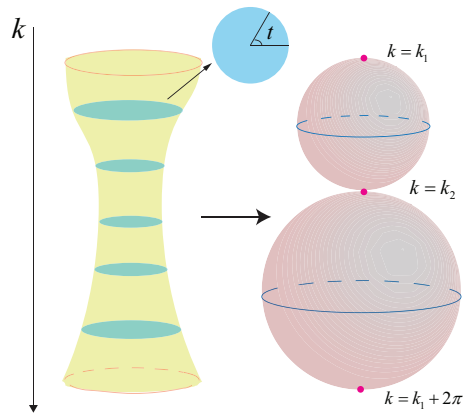}\\
  \caption{Scheme of the momentum-time manifold. In the left figure, for any fixed momentum k, the cross section can be viewed as a circle $S^1$ with the azimuthal angle represents the time t. After gluing $k=0$ and $k=2\pi$ (saffron circles), the topology of momentum-time manifold become $T^2$. If there are two fixed points $k=k_1$ and $k_2$, the corresponding circle contract to a point, then the momentum-time manifold can be reduced to a series of spheres $S^2$.}\label{Fig1}
\end{figure}

{\it General definition of dynamical topological invariant.-}
To unveil the dynamical properties after a sudden quench, we first study the manifold composed of momentum and time. In general, the 1D Brillouin zone (BZ) has the topology $S^1$. Furthermore, it can be seen from Eq. (\ref{Eq5}) that the time evolution of density matrices has a periodicity $\frac{\pi}{|\mathbf{d}^f|}$ for each momentum k. After rescaling the time $t^{\prime}=\frac{t}{|\mathbf{d}^f|}$, the topology of time can be viewed as $S^1$, hence the total momentum-time manifold has the topology $T^2$, as shown in the left column of Fig.\ref{Fig1}.

However, the momentum-time manifold is more complicated for topologically nontrivial 1D system. Suppose that there exist some momenta, at which 
$\mathbf{d}^i(k)$ are parallel or anti-parallel to $\mathbf{d}^f(k)$,
then the Bloch vectors $\hat{\mathbf{d}}(k,t)$ shall keep staying at the initial points $\mathbf{d}^i(k)$ according to Eq.(\ref{Eq5}), and the corresponding eigenvectors  $|\psi_{-}(k)\rangle$ will not evolve with time apart from a global phase. For topologically nontrivial systems, there must exist some ``fixed points", which are ensured by symmetries protecting the 1D topological state, as shall be further clarified later in specific cases belonging to different symmetry classes.
In the right column in Fig. \ref{Fig1}, we assume $k=k_1$ and $k_2$ are fixed points, at which the time axis can be contracted to a point as the Bloch vectors keep still during the time evolution. Therefore, the topology of the momentum-time manifold is reduced to two spheres \cite{NOTE2}. In general, if there are $N$ fixed points in a BZ, the momentum-time manifold $T^2$ can be reduced to $N$ submanifolds, each of them has the topology $S^2$.

Now we naturally have a map from each momentum-time submanifold to the Bloch vector $\hat{\mathbf{d}}(k,t)$, which is $S^2\rightarrow S^2$, and then we can define a Chern number
\begin{equation}\label{Eq6}
C_{dyn}^{m} = \frac{1}{4\pi}\int_{k_m}^{k_{m+1}}dk\int_{0}^{\pi}dt^{\prime} (\hat{\mathbf{d}}\times\partial_{t^{\prime}} \hat{\mathbf{d}})\cdot\partial_k , \hat{\mathbf{d}}
\end{equation}
where $m=1,2,...,N$ denotes the $m$th submanifold and $k_{m}$ denotes the $m$th fixed point, besides, $k_{N+1}=k_{1}+2\pi$ is the same point modulo $2\pi$. Hence the integral is an integer and it measures how many times the Bloch vectors cover the unit sphere.

By a straightforward calculation \cite{SUP}, the dynamical topological invariant can be written as
\begin{equation}\label{Eq7}
C_{dyn}^{m}=\frac{1}{2}(\cos\theta_{k=k_{m}}-\cos\theta_{k=k_{m+1}}),
\end{equation}
where $\theta_k$ is the included angle between $\hat{\mathbf{d}}^i(k)$ and $\hat{\mathbf{d}}^f(k)$. This invariant only contains the information of fixed points $k=k_{m}$ and $k=k_{m+1}$, and depends only on the included angle of the initial and final Hamiltonian. Exchanging the initial and final parameters $\mathbf{d}^{i(f)}\rightarrow\mathbf{d}^{f(i)}$ shall not affect the dynamical topological invariant. On the other hand, the included angle $\theta_{k_{m}}$ can only take the value $0$ or $\pi$ as $\mathbf{d}^i(k)$ is parallel or anti-parallel to $\mathbf{d}^f(k)$, hence the dynamical topological invariant can only be $C_{dyn}^{m}=0$ for $\theta_{k_{m}} = \theta_{k_{m+1}}$, or $C_{dyn}^{m}=\pm1$ for $\theta_{k_{m}} = \theta_{k_{m+1}}+ \pi (\mod 2\pi)$.

For any $m\in1,2,...,N$, the mapping from the corresponding momentum-time submanifold to the Bloch vectors induces a dynamical topological invariant $C_{dyn}^{m}$, and $C_{dyn}^{m}=0$ ($\pm1$) indicates that the mapping is trivial (nontrivial). The number of nontrivial mappings are related to the topological properties of initial and final systems. In the following we shall unveil these relations by studying specific topological classes of the ten-fold way symmetry classification \cite{APS}. According to this classification, for 1D two-band systems with no internal spin degree, only three types, i.e., classes of BDI, AIII and D, are topologically nontrivial.

{\it Class BDI and AIII.-}
In 1D, systems of class BDI and AIII preserve chiral symmetry and hence the topology invariant are characterized by winding number, and
the relation between the number of nontrivial mappings $M$ and the winding numbers in equilibrium is given by:

{\bf Theorem 1} {\it In class BDI and AIII, the number of nontrivial mappings $M$ from momentum-time submanifolds to Bloch vectors has a lower bound $2|n^i-n^f|$, with $n^i$ and $n^f$ being the winding numbers of $H^i$ and $H^f$, respectively.}

See SM \cite{SUP} for proof and details. Note that $M$ can reflect the difference of winding numbers between initial and final Hamiltonians. If $n^i\ne n^f$, the nontrivial mappings can not be removed simultaneously under continuous deformation because the winding number in equilibrium is protected by symmetry.
In most of examples which we have met there are only two fixed points, it can be found from Eq. (\ref{Eq7}) that the dynamical topological invariants satisfy $C_{dyn}^{1}=-C_{dyn}^{2}$, then only the $m=1$ submanifold is needed to know, and we therefore have the following corollary from theorem 1:

{\bf Corollary 1} {\it Consider a Hamiltonian in class BDI and AIII, suppose that there are only two fixed points, then the dynamical topological invariant $C_{dyn}^{1}=\pm 1$ if $H^i$ and $H^f$ lie in different topological phases.}

From corollary 1 we find that the dynamical topological invariant $C_{dyn}^{1}$ is closely related to topological properties of both $H^i$ and $H^f$.
To see it clearly, next we study a concrete example by considering the famous SSH \cite{SSH} model, which belongs to the class BDI and is described by the Hamiltonian:
\begin{eqnarray}
H=\sum_i [(t+\delta)\hat{c}^{\dagger}_{A,i}\hat{c}_{B,i}+(t-\delta
)\hat{c}^{\dagger}_{A,i+1}\hat{c}_{B,i}]+h.c.,
\end{eqnarray}
where $\hat{c}^{\dagger}_{A(B),i}$ is the creation operator of fermion on the
$i$-th A (or B) sublattice.  After the Fourier
transformation $\hat{c}_{s,j}=\frac{1}{\sqrt{L}}\sum_k e^{ikj}\hat{c}_{k,s}$ with $s=A (B)$ and setting $t=1$,
the Hamiltonian can be written as
\begin{eqnarray}
H = \sum_{k} \psi^{\dagger}_{k} h(k) \psi_{k} , \label{H_k}
\end{eqnarray}
where
$\psi^{\dagger}_{k}=(\hat{c}^{\dagger}_{k,A},\hat{c}^{\dagger}_{k,B})$
and $h(k)=d_x\sigma_x+d_y\sigma_y$, with $d_x = (1+\delta)+(1-\delta)\cos k$ and $d_y = (1-\delta)\sin k$. For $\delta>0$, the half-filled system is topologically trivial, whereas for $\delta<0$ the system is topological. It can be seen for any $\delta^i$ and $\delta^f$, there are only two fixed points $k_1=0$ and $k_2=\pi$, corresponding to two high-symmetry points \cite{YC}, and the total Brilloin zone is reduced to two spheres denoted by $m=1,2$ with $k\in(0,\pi)$ and $k\in(\pi,2\pi)$, respectively.

Suppose that both $\delta^i$ and $\delta^f$ lie in the same phase, either topologically nontrivial or trivial, the Bloch vectors satisfy $\hat{\mathbf{d}}^i(0)=\hat{\mathbf{d}}^f(0)$ and $\hat{\mathbf{d}}^i(\pi)=\hat{\mathbf{d}}^f(\pi)$. According to Eq. (\ref{Eq7}), we have $C_{dyn}^1=C_{dyn}^2=0$, hence the number of nontrivial mappings is zero, which equals to difference of winding numbers between initial and final Hamiltonians. On the other hand, if $\delta^i$ and $\delta^f$ are in different phases, we find $\hat{\mathbf{d}}^i(0)=\hat{\mathbf{d}}^f(0)$ and $\hat{\mathbf{d}}^i(\pi)=-\hat{\mathbf{d}}^f(\pi)$. The dynamical topological invariants are $C_{dyn}^1=-1$ and $C_{dyn}^2=1$, indicating that both of mappings are topologically nontrivial. Our results show that the number of nontrivial mappings in the SSH model is $M=2|n^i-n^f|$, in accordance with theorem 1.

\begin{figure}[htbp]
  \centering
  \includegraphics[width=0.9\linewidth]{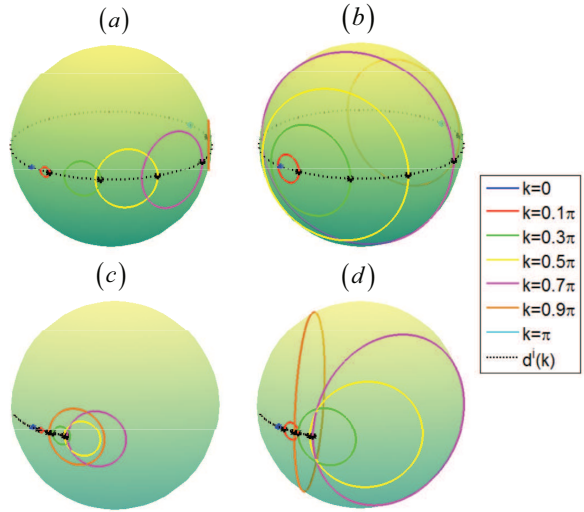}\\
  \caption{The evolution of Bloch vector for different momenta in a period, with (a)$\delta^i=-0.7$ and $\delta^f=-0.3$; (b)$\delta^i=-0.7$ and $\delta^f=0.3$; (c)$\delta^i=0.4$ and $\delta^f=0.2$; (d)$\delta^i=0.4$ and $\delta^f=-0.2$. The black dotted loop represents the distribution of initial Bloch vectors $\hat{\mathbf{d}}^i(k)$, for (a) (b) corresponding to topological cases and (c)(d) trivial cases. The black stars at the crossing points represent the initial points at $t=0$ of each loop.}\label{Fig2}
\end{figure}

Because there are only two fix points in the Hamiltonian, it is sufficient to study the submanifold $m=1$ from corollary 1. In order to get an intuitive understanding of the geometrical meaning of the dynamical Chern number, in Fig. \ref{Fig2} we show the evolution of Bloch vectors for different choices of initial and final parameters for the submanifold $m=1$. Each solid loop represents the trajectory of a definite mode in a period. After collecting all trajectories for $k\in(0,\pi)$ into the Bloch sphere, the topological properties can be directly obtained. In Fig. \ref{Fig2}(a) and (c), both $\delta^i$ and $\delta^f$ lie in the same phase, and the trajectories of different Bloch vectors cancel out each other, giving rise to the dynamical Chern number $C_{dyn}^1=0$; however, in Fig. \ref{Fig2}(b) and (d), $\delta^i$ and $\delta^f$ lie in different phases, and the trajectories for Bloch vectors of $k\in(0,\pi)$ cover the whole Bloch sphere, corresponding to $C_{dyn}^1=-1$.

{\it Class D.-}
In class D, the particle-hole symmetry constrains the direction of Bloch vector $\hat{\mathbf{d}}(k)$ lying on the z-axis at $k=0$ and $k=\pi$ \cite{AYK}, hence they are fixed points and the momentum-time submanifolds are reduced to two spheres denoted by $m=1,2$. The $Z_2$ topological invariant for static 1D systems in class D can be defined as $n=sgn(d_z(0)d_z(\pi))$,
where $n=-1$ for topological phase and $n=1$ for trivial phase. Similar to class BDI and AIII, we can prove \cite{SUP}:


{\bf Theorem 2} {\it Consider a Hamiltonian in class D, the dynamical topological invariant $C_{dyn}^{1}=0$ if $H^i$ and $H^f$ lie in the same phase, in contrast, $C_{dyn}^{1}=\pm 1$ if $H^i$ and $H^f$ lie in different phases.}

For the system of class D, since there exist solely two fixed points $k=0$ and $\pi$, protected by particle-hole symmetry, and thus we only need calculate the dynamical topological invariant of submanifold $m=1$.
A simple example of the D class system is the extended version of 1D Kitaev model described by Hamiltonian (\ref{Eq1}) with $d_0=0$, $d_x=\Delta_2\sin\phi\sin 2k$, $d_y=\Delta_1\sin k+\Delta_2\cos\phi\sin 2k$ and $d_z=-t_1\cos k-t_2\cos 2k+\mu$ \cite{AYK,LL}, where $t_1$ and $t_2$ represent the nearest-neighbor (NN) and next-nearest-neighbor (NNN) hopping amplitudes,  $\Delta_1$ and $\Delta_2$ the NN and NNN pairing parameters, $\phi$ denotes the phase difference of the two pairing parameters, and $\mu$ the chemical potential. For simplicity, we take $\mu=1$, $\Delta_1=t_1=0.5$, $\phi=\frac{\pi}{2}$ and $\Delta_2=t_2\equiv\Delta$. It was shown that the system is topological when $0.5<\Delta<1.5$, with Majorana fermions emerging at boundaries, whereas the system is trivial for $\Delta<0.5$ and $\Delta>1.5$. The fixed points $k=0$ and $\pi$ are protected by particle-hole symmetry, and explicitly we have $\hat{\mathbf{d}}(k=0)=(0,0,sgn(0.5-\Delta))$ and $\hat{\mathbf{d}}(k=\pi)=(0,0,sgn(1.5-\Delta))$. By using the formula (\ref{Eq7}), we can directly calculate the dynamical topological invariant.
Because both the interval $\Delta<0.5$ and $\Delta>1.5$ are topologically trivial, there are more choices for initial and final parameters. The dynamical topological invariant are shown in Table. \ref{Tab1}, indicating that nontrivial dynamical topological invariant appears only when $\Delta^i$ and $\Delta^f$ lie in topologically different phases.

\begin{figure}[htbp]
  \centering
  \includegraphics[width=0.9\linewidth]{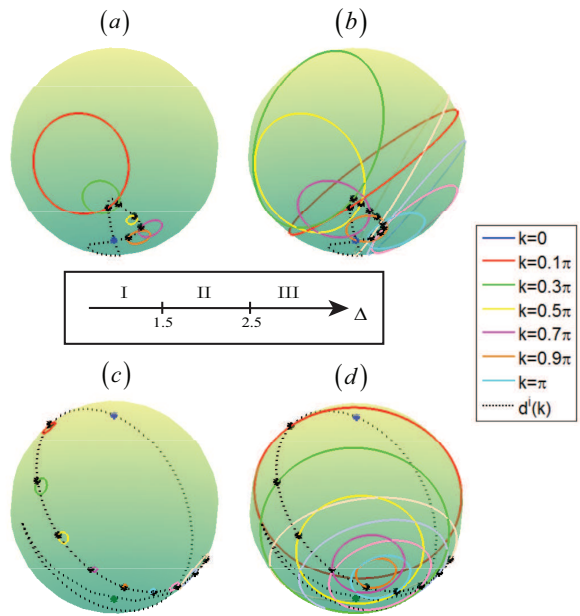}\\
  \caption{The evolution of Bloch vector for different momenta in a period, with (a)$\Delta^i=0.2$ and $\Delta^f=0.4$; (b)$\Delta^i=0.2$ and $\Delta^f=1.4$; (c)$\Delta^i=1.4$ and $\Delta^f=1.2$; (d)$\Delta^i=1.4$ and $\Delta^f=0.2$. The black dotted loop denotes the distribution of initial Bloch vectors $\hat{\mathbf{d}}^i(k)$, corresponding to trivial case in (a)(b) and topological case in (c)(d). The black stars at the crossing points represent the initial points at $t=0$ of each loop. The middle inset gives the phase diagram of the static system. }\label{Fig3}
\end{figure}

To get a geometrical understanding, in Fig. \ref{Fig3}, we show the evolution of Bloch vectors for different choices of $\Delta^i$ and $\Delta^f$s. In class D, though the initial states (black dotted curve) do not lie on an orthodrome, and the trajectories seem to be more complicated, we can always find the Bloch sphere being fully covered in (b) and (d) with $\Delta^i$ and $\Delta^f$ located in different phases, while all the trajectories cancel out each other in (a) and (c) with $\Delta^i$ and $\Delta^f$ located in the same phase.

\begin{table}
\centering
\begin{tabular}{|c|c|c|}
  \hline
  initial to final phases & $m=1$ & $m=2$ \\
  \hline
  I $\rightarrow$ I & 0 & 0 \\
  I $\rightarrow$ II & 1 & -1 \\
  I $\rightarrow$ III & 0 & 0 \\
  II $\rightarrow$ II & 0 & 0 \\
  II $\rightarrow$ III & -1 & 1 \\
  III $\rightarrow$ III & 0 & 0 \\
  \hline
\end{tabular}
\caption{Dynamical topological invariant for different initial and final phases. For convenience we label the interval $\Delta<0.5$ by I, $0.5<\Delta<1.5$ by II and $\Delta>1.5$ by III, hence the interval II is topological phase, the interval I and III are topological trivial phase, and $m=1,2$ labels the momentum-time submanifolds.}
\label{Tab1}
\end{table}

{\it A topologically trivial example.-}
As a comparison, we study a 1D tight binding model with alternating chemical potential $\mu$ and $-\mu$ on the A and B sublattice, which is described by Hamiltonian (\ref{H_k}) with $d_x=1+\cos k$, $d_y=\sin k$ and $d_z=\mu$. The system carries out a quantum phase transition at $\mu=0$, which is just the gap-closing point. This model describes a topologically trivial system as it always has no nontrivial Berry phase \cite{LH-2015}. We find that there exists only an accidental fixed point $k=\pi$, and thus the dynamical topological invariant should be calculated in the whole BZ.  According to the formula (\ref{Eq7}), the dynamical topological invariant $C_{dyn}^{1}=0$ for arbitrary $\mu^i$ and $\mu^f$s, suggesting that no nontrivial mappings from momentum-time manifold to Bloch vectors exist in this model.



Finally, we point out that the dynamical topological invariant is in principe measurable in current cold atom experiments. If the energy band of $H^f$ is flat band, i.e. SSH model with $\delta^f=\pm 1$, all the momenta have the same period. With the technique of Bloch state tomography \cite{CW,JJG,AE}, the evolution of the state in BZ can be observed in a period. After collecting the trajectories in a definite momentum-time submanifold $m$, the dynamical topological invariant can be measured directly.

{\it Summary.-}
In summary, we have clarified how to properly define a dynamical topological invariant for the general 1D two-band insulator system performed by a quantum quench. After showing that the momentum-time manifold can be reduced to a series of spheres, a dynamical topological invariant can be defined for the mapping from each of the sphere to the Bloch vectors. Then we analyze the intrinsic relation between the dynamical topological invariant and the topological invariant of $H^i$ and $H^f$ in equilibrium. We also give some visualized examples to show our results. Finally, we point out that the dynamical topological invariant is experimentally measurable.

\begin{acknowledgments}
The work is supported by the National Key Research and Development Program of China (2016YFA0300600), NSFC under Grants No. 11425419, No. 11374354 and No. 11174360, and the Strategic Priority Research Program (B) of the Chinese Academy of Sciences  (No. XDB07020000).
\end{acknowledgments}

\onecolumngrid
\newpage
\section{Supplementary Material
}
\subsection{Evolution of density matrices}
In this section, we calculate the evolution of density matrices. From the Liouville-von Neumann equation,
\begin{equation}\label{S1}
i\frac{\partial\rho(k,t)}{\partial t}=[h^f(k),\rho(k,t)],
\end{equation}
it can be found that the solution is
\begin{equation}\label{S2}
\rho(k,t)=e^{-ih^f(k)t}\rho(k,0)e^{ih^f(k)t},
\end{equation}
where $\rho(k,0)=\rho^i(k)=\frac{1}{2}[1-\hat{\mathbf{d}}^i(k)\cdot\boldsymbol{\sigma}]$.
The density matrix can be generally written as $\rho(k,t)=\frac{1}{2}[1-\hat{\boldsymbol{d}}(k,t)\cdot\boldsymbol{\sigma}]$.
By using the following relation
\begin{gather}\label{S3}
  e^{i\mathbf{a}\cdot\boldsymbol{\sigma}}= \cos |\mathbf{a}|+i(\hat{\mathbf{a}}\cdot\boldsymbol{\sigma}) \sin |\mathbf{a}| \nonumber\\
  (\boldsymbol{\sigma}\cdot\mathbf{a}) (\boldsymbol{\sigma}\cdot\mathbf{b})= \mathbf{a}\cdot\mathbf{b}+i(\mathbf{a}\times\mathbf{b})\cdot\boldsymbol{\sigma},
\end{gather}
and after a straightforward calculation, we can derive
\begin{equation}\label{S4}
\hat{\mathbf{d}}(k,t) = \hat{\mathbf{d}}^i\cos(2|\mathbf{d}^f| t) + 2\hat{\mathbf{d}}^f(\hat{\mathbf{d}}^i\cdot\hat{\mathbf{d}}^f)\sin^2(|\mathbf{d}^f| t) + \hat{\mathbf{d}}^i\times\hat{\mathbf{d}}^f\sin(2|\mathbf{d}^f| t).
\end{equation}

\subsection{Calculation of dynamical topological invariant}
\begin{figure}[htbp]
  \centering
  \includegraphics[width=0.6\linewidth]{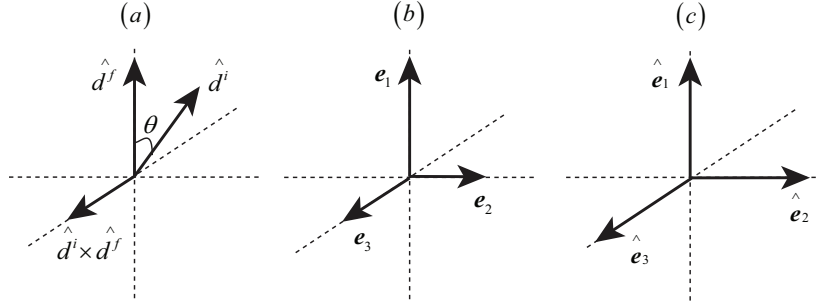}\\
  \caption{Scheme of Bloch vectors, we show the three group of bases where we have used in our calculation.}\label{FigS1}
\end{figure}

To calculate the dynamical Chern number, we should rewrite Eq. (\ref{S4}) as the following form:
\begin{equation}\label{S5}
\hat{\mathbf{d}}(k,t) = \mathbf{e}_1 + \mathbf{e}_2\cos(2\epsilon_f t) + \mathbf{e}_3\sin(2\epsilon_f t),
\end{equation}
with
\begin{equation}\label{S6}
\mathbf{e}_1 = \hat{\mathbf{d}}^f(\hat{\mathbf{d}}^i\cdot\hat{\mathbf{d}}^f),~~~~~~~~
\mathbf{e}_2 = \hat{\mathbf{d}}^i-\hat{\mathbf{d}}^f(\hat{\mathbf{d}}^i\cdot\hat{\mathbf{d}}^f),~~~~~~~~
\mathbf{e}_3 = \hat{\mathbf{d}}^i\times\hat{\mathbf{d}}^f,
\end{equation}
where the new bases $(\mathbf{e}_1,\mathbf{e}_2,\mathbf{e}_3)$ are orthogonal as shown in Fig. \ref{FigS1}(b). With the new bases, and substituting Eq. (\ref{S5}) into Eq. (\ref{Eq6}), we can derive
\begin{equation}\label{S7}
C_{dyn} = -\frac{1}{4}\int_{k_m}^{k_{m+1}}dk(2\mathbf{e}_3\times\mathbf{e}_2\cdot\partial_k\mathbf{e}_1
- \partial_k\mathbf{e}_3\times\mathbf{e}_2\cdot\mathbf{e}_1 - \mathbf{e}_3\times\partial_k\mathbf{e}_2\cdot\mathbf{e}_1).
\end{equation}
Using the condition that angle between $\hat{\mathbf{d}}^i(k)$ and $\hat{\mathbf{d}}^f(k)$ is $\theta(k)$, we can define the normalized bases
\begin{equation}\label{S8}
\hat{\mathbf{e}}_1 = \frac{\mathbf{e}_1}{\cos\theta}~~~~~~~~~~
\hat{\mathbf{e}}_2 = \frac{\mathbf{e}_2}{\sin\theta}~~~~~~~~~~
\hat{\mathbf{e}}_3 = \frac{\mathbf{e}_3}{\sin\theta},
\end{equation}
which is shown in Fig. \ref{FigS1}(c), then Eq. (\ref{S7}) can be simplified by
\begin{equation}\label{S9}
C_{dyn} = -\int_{k_{m}}^{k_{m+1}}dk [-\frac{1}{2}\sin\theta\frac{d\theta}{dk} + \frac{1}{4}\sin^2\theta\cos\theta (2\hat{\mathbf{e}}_3\times\hat{\mathbf{e}}_2\cdot\partial_k\hat{\mathbf{e}}_1 - \partial_k\hat{\mathbf{e}}_3\times\hat{\mathbf{e}}_2\cdot\hat{\mathbf{e}}_1 - \hat{\mathbf{e}}_3\times\partial_k\hat{\mathbf{e}}_2\cdot\hat{\mathbf{e}}_1)].
\end{equation}
Furthermore, because the derivative of the unit vector is perpendicular to itself,
\begin{equation}\label{S10}
\hat{\mathbf{e}}_\alpha\cdot\partial_k\hat{\mathbf{e}}_\alpha=0,
\end{equation}
with $\alpha=1,2,3$, we have
\begin{equation}\label{S11}
\partial_k\hat{\mathbf{e}}_\alpha\times\hat{\mathbf{e}}_\beta\cdot\hat{\mathbf{e}}_\gamma=0,
\end{equation}
with $\alpha,\beta,\gamma$ being a permutation of $(1,2,3)$. Finally we derive
\begin{equation}\label{S12}
C_{dyn}=\frac{1}{2}(\cos\theta_{k=k_{m}}-\cos\theta_{k=k_{m+1}}).
\end{equation}
\subsection{Properties of fixed points and dynamical topological invariant}
In this section, we give the proof of theorem 1 and 2, which give the intrinsic relations between the fixed points and dynamical topological invariant.

{\bf Theorem 1} {\it In class BDI and AIII, the number of nontrivial mappings $M$ from momentum-time submanifolds to Bloch vectors has a lower bound $2|n^i-n^f|$, with $n^f$ and $n^f$ being the winding numbers of $H^i$ and $H^f$, respectively.}

{\bf Proof} Consider a Hamiltonian $H=\sum_{k}\psi^{\dagger}_{k}h(k)\psi_{k}$ with
$h(k)=\mathbf{d}\cdot\boldsymbol{\sigma}$ in class BDI or AIII \cite{APSs,SRYs}, the chiral symmetry requires $Sh(k)S^{-1}=-h(k)$, and the chiral operator $S=e^{i\frac{\alpha}{2}}e^{i\frac{\boldsymbol{\sigma}}{2}\cdot\mathbf{a}}$ is a two-dimensional unitary matrix. From the identity
\begin{equation}\label{S13}
e^{i\frac{\boldsymbol{\sigma}}{2}\cdot\mathbf{a}}(\mathbf{d}\cdot\boldsymbol{\sigma})e^{-i\frac{\boldsymbol{\sigma}}{2}\cdot\mathbf{a}} = \mathbf{d}\cdot\boldsymbol{\sigma}\cos|\mathbf{a}| + \mathbf{d}\cdot(\hat{\mathbf{a}}\times\boldsymbol{\sigma})\sin|\mathbf{a}| + (\mathbf{d}\cdot\hat{\mathbf{a}})(\hat{\mathbf{a}}\cdot\boldsymbol{\sigma})[1-\cos|\mathbf{a}|]
\end{equation}
we see that the chiral symmetry leads to $|\mathbf{a}|=\pi$ and $\mathbf{d}\cdot\hat{\mathbf{a}}=0$. Hence the Bloch vector $\mathbf{d}(k)$ is perpendicular to the vector $\mathbf{a}$ and lies in an orthodrome for different $k$. Since the topological invariant of the static system is characterized by a winding number, we only need to consider the normalized Bloch vector $\hat{\mathbf{d}}(k)$ without changing the topological properties.

Then we can parameterize the Bloch vector $\hat{\mathbf{d}}(k)$ by a polar angle $\phi(k)$, and the winding number can be written as
\begin{equation}\label{S14}
n=\oint dk \frac{\partial\phi(k)}{\partial k}.
\end{equation}
Suppose that $\phi^i(k)$ and $\phi^f(k)$ are the polar angle of $\hat{\mathbf{d}}^i(k)$ and $\hat{\mathbf{d}}^f(k)$, respectively, then we have  $n^i=\oint dk \frac{\partial\phi^i(k)}{\partial k}$ and $n^f=\oint dk \frac{\partial\phi^f(k)}{\partial k}$.
For convenience, we define a new Bloch vector $\hat{\mathbf{d}}^0(k)$, which has the polar angle $\theta(k)=\phi^i(k)-\phi^f(k)$,
and the winding number of $\hat{\mathbf{d}}^0(k)$ is $n^i-n^f$.
At the fixed points, we have $\hat{\mathbf{d}}^i(k)=\pm\hat{\mathbf{d}}^f(k)$, which gives rise to $\phi^i(k)=\phi^f(k)$ or $\phi^i(k)=\phi^f(k)+\pi$. Correspondingly, the polar angle of the Bloch vector $\hat{\mathbf{d}}^0(k)$ at fixed points fulfills $\theta(k)=0$ or $\pi$(mod $2\pi$).
Equivalently, seeking the number of mapping from the momentum-time submanifold to the Bloch sphere equals to counting the number of $\theta(k)=0$ or $\pi$(mod $2\pi$) when $k$ goes round the BZ.
Because the winding number of $\hat{\mathbf{d}}^0(k)$ is $n^i-n^f$, then there will be at least $|n^i-n^f|$ fixed points satisfying $\theta(k)=0$ and
$|n^i-n^f|$ fixed points satisfying $\theta(k)=\pi$, i.e., altogether there exist at least $2|n^i-n^f|$ fixed points.

First we consider the number of fixed points is $N=2|n^i-n^f|$, and denote the fixed points satisfying $\theta(k)=0$ by $k=k^a_m$, and the fixed points satisfying $\theta(k)=\pi$ by $k=k^b_m$, where $m=1,2,...,|n^i-n^f|$ is the label of fixed points. It can be seen that the two types of fixed points emerge staggeredly as $k$ increases, i.e., in a sequence ($k^a_1,k^b_1,k^a_2,k^b_2,\dots,k^a_{\frac{N}{2}},k^b_{\frac{N}{2}}$). Then from Eq. \ref{S12} all the dynamical topological invariants are nontrivial.

If the number of fixed points is larger than $2|n^i-n^f|$, we denote the additional fixed points by $k^c$ which may satisfy either $\theta(k^c)=0$ or $\theta(k^c)=\pi$. As $k$ increases, the fixed points  emerge in a sequence ($k^a_1,k^b_1,k^a_2,k^c_1,k^c_2,k^b_2,\dots,k^a_{\frac{N}{2}},k^c_n,k^b_{\frac{N}{2}}$), where $k^c$s may lie in anywhere among the original sequence. As a result, the number of fixed points or the momentum-time submanifolds will increase, and the nontrivial Chern number may increase.

{\bf Corollary 1} {\it Consider a Hamiltonian in class BDI and AIII, and suppose that there are only two fixed points, then the dynamical topological invariant $C_{dyn}^{1}=\pm 1$ if $H^i$ and $H^f$ lie in different topological phases.}

{\bf Proof} Suppose that there are only two fixed points $k_1$ and $k_2$, then there are total $N=2$ momentum-time submanifolds. First we can see from theorem 1 that the number of nontrivial mappings is $N\ge M\ge2|n^i-n^f|$, hence the static Hamiltonian can not have winding number larger than $1$. Then if $H^i$ and $H^f$ lie in different phases, i.e. $|n^i-n^f|=1$, the dynamical topological invariant $C_{dyn}^{1}=-C_{dyn}^{2}=\pm 1$.

{\bf Theorem 2} {\it Consider a Hamiltonian in class D, the dynamical topological invariant $C_{dyn}^{1}=0$ if $H^i$ and $H^f$ lie in the same phase, in contrast, $C_{dyn}^{1}=\pm 1$ if $H^i$ and $H^f$ lie in different phases.}

{\bf Proof} In class D, the fixed points are $k=0,\pi$ protected by the particle-hole symmetry, and the Bloch vectors are constrained on the south and north poles. The $Z_2$ topological invariant can be defined as $n=sgn(d_z(0)d_z(\pi))$ \cite{AYKs}, where $n=-1$ for the topological phase and $n=1$ for the trivial phase.

If $H^i$ and $H^f$ lie in the same phase, we find
\begin{equation}\label{S15}
n^in^f=sgn(d^i_z(0)d^i_z(\pi)d^f_z(0)d^f_z(\pi))=1 .
\end{equation}
Equivalently, it can be rewritten as
\begin{equation}\label{S16}
sgn(d^i_z(0)d^f_z(0))=sgn(d^i_z(\pi)d^f_z(\pi)),
\end{equation}
which implies both included angles $\theta_{k=0}=\theta_{k=\pi}=0$ or $\pi$. Consequently, the dynamical topological invariant $C_{dyn}^{1}=0$.

On the other hand, if $H^i$ and $H^f$ lie in different phases, we can see
\begin{equation}\label{S17}
n^in^f=sgn(d^i_z(0)d^i_z(\pi)d^f_z(0)d^f_z(\pi))=-1,
\end{equation}
and it can be rewritten as
\begin{equation}\label{S18}
sgn(d^i_z(0)d^f_z(0))=-sgn(d^i_z(\pi)d^f_z(\pi)),
\end{equation}
which gives rise to $\theta_{k=0}=0$ and $\theta_{k=\pi}=\pi$, or $\theta_{k=0}=\pi$ and $\theta_{k=\pi}=0$.
If $\theta_{k=0}=0$ and $\theta_{k=\pi}=\pi$, the dynamical topological invariant $C_{dyn}^{1}=-1$; in contrary, if $\theta_{k=0}=\pi$ and $\theta_{k=\pi}=0$, the dynamical topological invariant $C_{dyn}^{1}=1$.


\begin{references}
\bibitem{NOTE2} The crossing point belongs to two submanifolds at the same time, see the pink points($k=k_1$ and $k_2$) in Fig. \ref{Fig1}, so the total manifold is not the direct sum of each submanifold, but it does not impact the following definition of dynamical topological invariant.
\end{references}
\end{document}